\documentclass[
 reprint,
 amsmath,amssymb,
 aps,
]{revtex4-2}
\usepackage[utf8]{inputenc}
\usepackage{newunicodechar}
\newunicodechar{−}{-}
\usepackage{graphicx}
\usepackage{dcolumn}
\usepackage{bm}
\usepackage{siunitx}
\usepackage[hidelinks]{hyperref}
\usepackage{gnuplottex} 
\usepackage[hidelinks]{hyperref}
\usepackage{hyphenat}

\PassOptionsToPackage{hyphens}{url}
\usepackage{hyperref}
\hypersetup{breaklinks=true}

\usepackage{subfig}
\usepackage{ragged2e}
\usepackage{caption}

\DeclareCaptionFormat{apsblock}{\justifying #1#2#3}
\newlength{\subcolumnwidth}

\newcommand{\nextsubcolumn}[1][]{
  \cr\noalign{\hfill}
  \if\relax\detokenize{#1}\relax\else\hsize=#1\setlength{\subcolumnwidth}{\hsize}\fi
}

\begin{document}

\preprint{APS/123-QED}

\title[]{The microscopic origin of droplet line tension}

\author{Franziska Aurbach}
\affiliation{ 
Institute of Applied Materials-Microstructure Modelling and Simulation, 
Karlsruhe Institute of Technology, 
Stra{\ss}e am Forum 7, 
76131 Karlsruhe, Germany
}
\affiliation{ Institute of Nanotechnology, 
Karlsruhe Institute of Technology, 
Hermann-von-Helmholtz-Platz 1,
76344 Eggenstein-Leopoldshafen, Germany}
\author{Fei Wang}
\email{fei.wang@kit.edu}
\affiliation{ 
Institute of Applied Materials-Microstructure Modelling and Simulation, 
Karlsruhe Institute of Technology, 
Stra{\ss}e am Forum 7, 
76131 Karlsruhe, Germany
}
\affiliation{ Institute of Nanotechnology, 
Karlsruhe Institute of Technology, 
Hermann-von-Helmholtz-Platz 1,
76344 Eggenstein-Leopoldshafen, Germany}

\author{Britta Nestler}
\affiliation{ 
Institute of Applied Materials-Microstructure Modelling and Simulation, 
Karlsruhe Institute of Technology, 
Stra{\ss}e am Forum 7, 
76131 Karlsruhe, Germany
}
\affiliation{ Institute of Nanotechnology, 
Karlsruhe Institute of Technology, 
Hermann-von-Helmholtz-Platz 1,
76344 Eggenstein-Leopoldshafen, Germany}
\affiliation{
Institute of Digital Materials Science, Karlsruhe University of Applied Sciences, 
Moltkestra{\ss}e 30, 
76133 Karlsruhe, Germany
}

\date{\today}

\begin{abstract}
The size dependence of the equilibrium droplet contact angle is governed by line tension. In this work, we identify a contribution to line tension arising from gravitational effects and pressure-induced changes in volume-fraction-dependent interfacial tensions within an adsorption layer. This mechanism addresses a multiscale problem of line tension in droplets ranging from nanometric to millimetric sizes that change sign and span several orders of magnitude, in agreement with experimental and simulation results. The sign of the apparent line tension is controlled by surface wettability, the initial volume fraction in the adsorption layer, and the droplet size, which also strongly influences its magnitude. Our results provide a unified physical interpretation of the experimentally observed variability in both the sign and magnitude of line tensions.
\end{abstract}

\maketitle

\begin{figure}
    \centering
    \includegraphics[width=\linewidth]{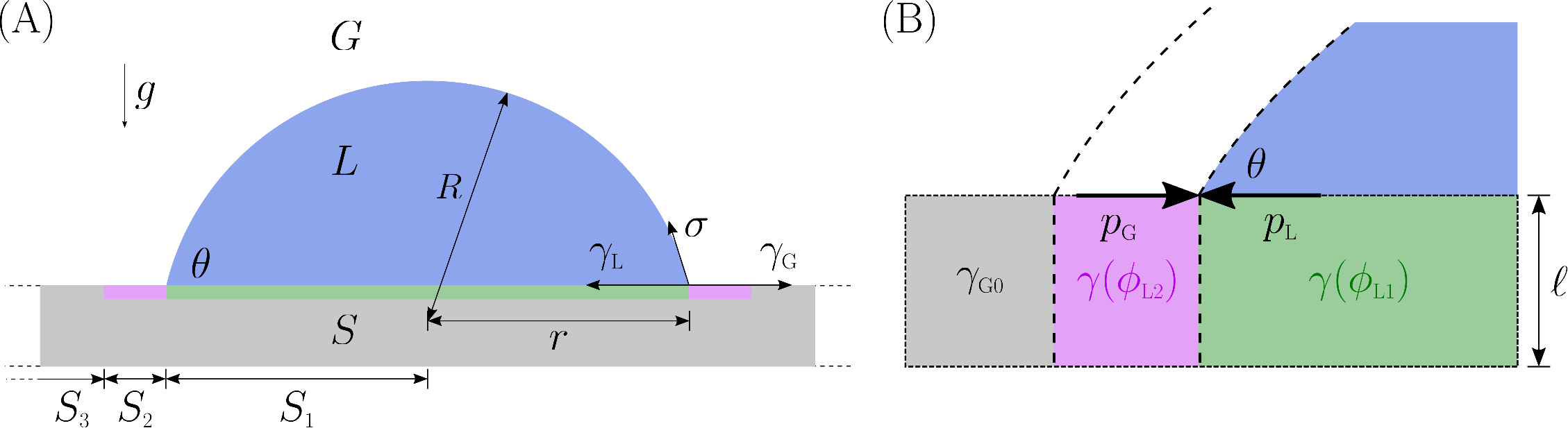}
    \caption{(A) Schematic cross-sectional representation of the liquid droplet (L, blue) on a solid substrate (S, gray) in a gas phase (G, white), illustrating the contact angle $\theta$, the base radius $r$, the cap radius $R$, and the interfacial tensions $\sigma$, $\gamma _G$, and $\gamma _L$. The colored layer represents the adsorption layer on the solid, which modifies the interfacial tensions. The wetted area, the local droplet penetration area and the exterior solid-gas interfacial area are represented by $S_1$, $S_2$, and $S_3$, respectively. (B) Enlarged view at the triple line and the adsorption layer of thickness $\ell$ showing the liquid volume fraction $\phi_{L1}$ and $\phi_{L2}$ dependent interfacial tensions $\gamma(\phi_{L1})$ (green) and $\gamma(\phi_{L2})$ (pink) in the vicinity of the droplet, which are affected by the liquid pressure $p_L$ and the gas pressure $p_G$, respectively. Outside the penetration area, the interfacial tension $\gamma_{G0}$ is considered to be constant (gray).
    }
    \label{figures/fig1.eps}
\end{figure}

\textit{Introduction ---} It has been demonstrated through numerous experiments that the equilibrium contact angle depends on the droplet size. This phenomenon cannot be explained within the framework of Young's law~\cite{youngIIIEssayCohesion1805}, thereby necessitating the introduction of line tension.
Line tension has been observed to not be a constant material parameter~\cite{iwamatsuGeneralizedYoungsEquation2018,kanducGeneralizedLineTension2018,klauserLineTensionDrop2022,zhangCriticalAssessmentLine2018,zhaoResolvingApparentLine2019}, but instead to span several orders of magnitude~\cite{amirfazliStatusThreephaseLine2004,marmurLineTensionIntrinsic1997}, while exhibiting both positive and negative signs in experimental and simulation results~\cite{zhaoResolvingApparentLine2019}.
The sign and magnitude of line tension are of particular relevance in the design and analysis of systems such as microfluidic devices, dropwise condensation~\cite{amirfazliMeasurementsLineTension2000} for heat transfer~\cite{amirfazliMeasurementsLineTension2000,amirfazliStatusThreephaseLine2004}, and heterogeneous nucleation~\cite{amirfazliLineTensionMeasurements1998,lawLineTensionIts2017}.
Despite extensive investigations into the physical origins of line tension, significant disagreement persists between existing theories and the line tension measured in experiments and simulations. A consensus on the physical origins underlying various signs and magnitudes of the measured line tension has, therefore, not yet been reached. 
Existing theories include several physical effects that are significant across different droplet sizes. At sub-nanometer atomic scales, line tension is attributed to poorly quantified direct atomic interactions~\cite{iwamatsuGeneralizedYoungsEquation2018}. At nanometric scales, the potential energy resulting from the attractive atomic forces between the liquid and the solid is modeled by the Lennard-Jones potential, with a liquid density that varies as a function of the distance from the surface, which requires a more complex numerical solution method~\cite{tanBodyForcesDrive2023}. In the case of millimetric droplets, the gravitational effect, which yields positive line tension, becomes predominant~\cite{iwamatsuGeneralizedYoungsEquation2018,tanBodyForcesDrive2023,lawLineTensionIts2017}. 

In this Letter, we present a unified theory across six orders of droplet size establishing a connection between line tension and the pressure effect on composition dependent interfacial tensions in an adsorption layer~\cite{wangWettingPhenomenaLine2024}. Our theory thus establishes a direct link between line tension and contact angle hysteresis on smooth substrates~\cite{wangWettingContactAngleHysteresis2024}. We show that two metastable adsorption states govern the sign and magnitude of the apparent line tension. When combined with gravitational effects, this approach enables the prediction of both the sign and the magnitude of line tension as functions of droplet size and surface wettability.

\textit{Model description ---} We consider a sessile droplet of volume $V$ on an ideally smooth and chemically homogeneous substrate, forming an apparent contact angle $\theta$ and a base radius $r$ (\autoref{figures/fig1.eps} (A)). 
The interfacial tensions of the liquid-gas, solid-gas, and solid-liquid phases are denoted by $\sigma$, $\gamma_G$, and $\gamma_L$, respectively. 
According to Young's law~\cite{youngIIIEssayCohesion1805}, the equilibrium contact angle satisfies $\cos \theta_0=(\gamma_G-\gamma_L)/\sigma$, which expresses the mechanical balance of interfacial tensions at the three-phase contact line.
In contrast, the modified Young's equation~\cite{BoruvkaGeneralization1977} incorporates size-dependent corrections to the apparent contact angle,
\begin{align}
    \cos \theta=\cos \theta_0-\frac{\tau}{\sigma r}. \label{mod-YL}
\end{align}
To elucidate the physical origin of the line tension $\tau$, we characterize the partial penetration of the liquid phase into the adsorption layer, thereby determining the dependence of the interfacial tensions on the droplet radius arising from pressure differences and yielding line tension.
In contrast to the common assumption of constant interfacial tensions, we treat $\gamma_L=\gamma(\phi_{L1},\phi_{G1})$ and $\gamma_G=\gamma(\phi_{L2},\phi_{G2})$ as functions of the average volume fractions of the liquid and gas species $\phi_{Li}=S_i^{-1}\int_{S_i}\phi_{L,\text{loc}}(\vec{x})\mathrm{d}S_i$ and $\phi_{Gi}=S_i^{-1}\int_{S_i}\phi_{G,\text{loc}}(\vec{x})\mathrm{d}S_i$ ($i=1,2$) over the wetted area $S_1$ and the local penetration area $S_2$ (\autoref{figures/fig1.eps} (B)), respectively. 
The local volume fractions of the liquid and gas species within the adsorption layer of thickness $\ell$ are denoted by $\phi_{L,\mathrm{loc}}(\vec{x})$ and $\phi_{G,\mathrm{loc}}(\vec{x})$, respectively.
In the exterior region $S_3$, the solid-gas interfacial tension $\gamma_{G0}$ is assumed to be constant. 
The interfacial tension is modeled as 
\begin{align} 
    \gamma(\phi_L,\phi_G)=\ell[\epsilon(\phi_L,\phi_G)+p(\phi_L,\phi_G)+\chi(\phi_L,\phi_G)]. \label{gamma}
\end{align} 
Here, $\epsilon$, $p$, and $\chi$ denote the internal, pressure, and van der Waals energy contributions, respectively (see end matter).
Our wall free energy formulation circumvents the difficulty of directly measuring the solid–fluid interfacial energies in experiments. It is constructed in analogy with the classical Gibbs free energy, 
$G=\epsilon+\chi+p−Ts$, incorporating contributions from internal energy, pressure energy, and the entropic term 
$Ts$. In the present treatment, however, the entropy contribution $s$ is neglected.
The wall free energy has units of $\unit{J/m^2}$, in contrast to the bulk Gibbs free energy, which has units of $\unit{J/m^3}$. This difference in dimensionality is reconciled, within the framework of mean-field theory, by introducing an average adsorption layer thickness $\ell$, which transforms the bulk free energy density into an effective surface free energy density. As will be demonstrated later, inclusion of the pressure energy in the wall free energy is essential for correctly capturing both the magnitude and the sign of the line tension. This constitutes the central contribution of the present work toward elucidating the fundamental mechanism underlying line tension.

In accordance with the constraint of the volume fractions, namely, $\phi_{L1} + \phi_{G1} = 1$ and $\phi_{L2} + \phi_{G2} = 1$, it can be concluded that only one independent variable remains. Consequently, the following simplified notation is employed: $\gamma(\phi_{L1})=\gamma(\phi_{L1},1-\phi_{L1})$ and $\gamma(\phi_{L2})=\gamma(\phi_{L2},1-\phi_{L2})$. 

The total free energy of the system is given by 
\begin{align} 
    E=\sigma A-\Delta \gamma S_1+\gamma(\phi_{L2}) (S_1+S_2)+\gamma_{G0}S_3+E_g. \label{energy}
\end{align} 
Here, the first three terms represent the interfacial energies, $E_g$ denotes the gravitational energy, and $A$ is the area of the droplet cap. The difference in interfacial tensions is given by $\Delta \gamma=\gamma(\phi_{L2})-\gamma(\phi_{L1})$.

\begin{figure*}[t]
    \centering
    \includegraphics[width=\linewidth]{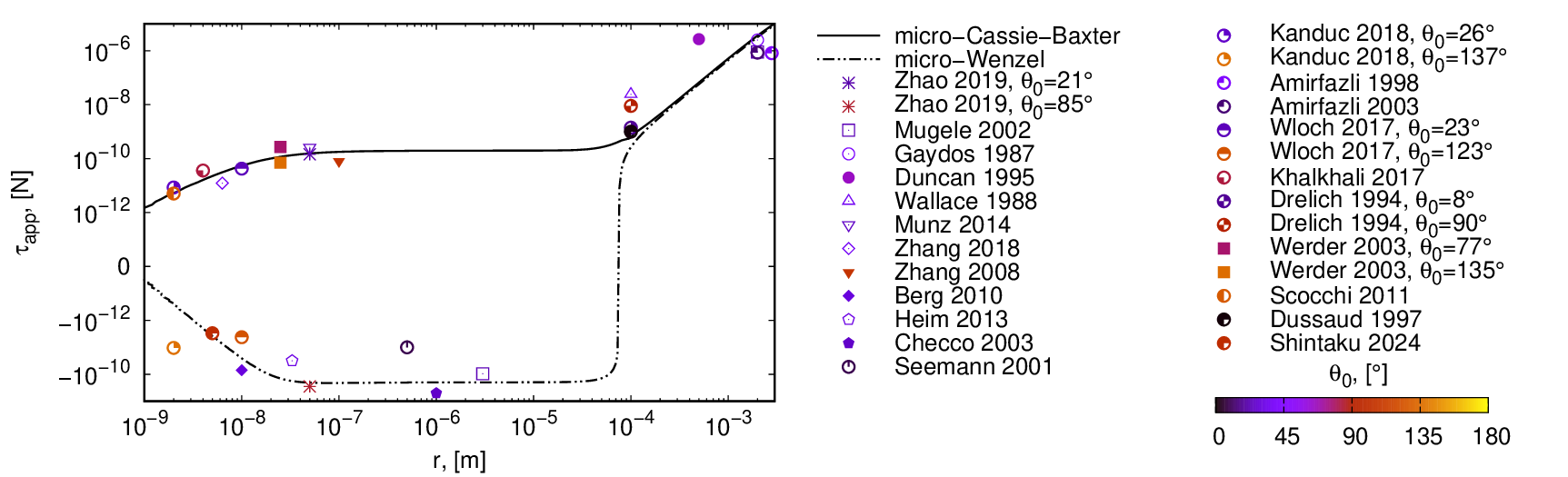}
    \caption{Line tension $\tau$ [N] as a function of the base radius $r$ [m] with $\theta_0=40^\circ$, $\Delta\rho=\qty{998}{kg\per m^{3}}$, $g=\qty{9.81}{m\per s^2}$, $\sigma=\qty{0.055}{N\per m}$, and $\ell=\qty{5}{nm}$, in comparison with previous experiments \cite{zhaoResolvingApparentLine2019,mugeleCapillarityNanoscaleAFM2002,gaydosDependenceContactAngles1987a,duncanCorrelationLineTension1995,wallaceLineTensionSessile1988a,munzSizeDependenceShape2014,zhangInterfacialOilDroplets2008,bergImpactNegativeLine2010,heimMeasurementLineTension2013,checcoNonlinearDependenceContact2003,seemannPolystyreneNanodroplets2001,amirfazliLineTensionMeasurements1998,drelicheffectSolidSurface1994,werderWaterCarbonInteractionUse2003,DussaudLineTension1996,AmirfazliDetermination2003} 
    and simulations \cite{wlochNewForcefieldWater2017,kanducGeneralizedLineTension2018,zhangCriticalAssessmentLine2018,scocchiWettingContactlineEffects2011,khalkhaliWettingNanoscaleMolecular2017,ShintakuMeasuring2024}, where Young's contact angle, $\theta_0$, is indicated by the color of each point.}
    \label{figures/r_tau_Zhao_mult.eps}
\end{figure*}

In this work, we use the spherical cap model for simplicity.
For millimetric droplets, the Bond number is of order unity; consequently, the ellipsoidal cap model would provide a more accurate representation of the droplet shape here~\cite{wangWettingPhenomenaLine2024,aurbachGravitationalEffectEquilibrium2025,whymanOblateSpheroidModel2009}, but is not considered in the present study.

In the spherical cap model, the interfacial areas make Eq.~\eqref{energy} to
\begin{align}
    E(\theta,R(\theta),\Delta\gamma)=\sigma 2\pi& R(\theta)^2(1-\cos \theta)\nonumber\\&-\Delta \gamma \pi R(\theta)^2\sin ^2\theta+ E_g(\theta), \label{eq-energy}
\end{align}
with $E_g(\theta)=\Delta\rho gV\overline{z}(\theta)$ and where $R$, $g$, and $\overline{z}$ denote the radius of the spherical cap, the gravitational acceleration, and the height of the droplet centroid relative to the solid substrate, respectively. The density difference between the liquid and gas phases is denoted by $\Delta \rho$.

We use the Young-Laplace equation to relate the pressure of the liquid phase $p_L$ to that of the gas phase $p_G$,
\begin{align}
    p_L=p_G+\frac{2\sigma}{R}.
\end{align}
The pressure difference influence $p_L-p_G$ alters the composition of the adsorption layer, thereby inducing a size dependence of the surface tension and a corresponding modification of the energy function.
By defining $\Delta \gamma_0(\phi_{L1},\phi_{L2}):=\gamma_0(\phi_{L2})-\gamma_0(\phi_{L1})$, we obtain a size-dependent expression
\begin{align}
    \Delta \gamma(\phi_{L1},\phi_{L2}) = \Delta \gamma _0(\phi_{L1},\phi_{L2}) +2\ell \Delta \phi \sigma /R, \label{Delta_gamma}
\end{align}
where $\Delta \phi=\phi_{L2}-\phi_{L1}$. The intrinsic part $\gamma_0(\phi_L,1-\phi_L)$ includes all terms in Eq.~\eqref{gamma} except the pressure term, leading to a reference Young's contact angle $\cos \theta_0=[\gamma_0(\phi_{L2})-\gamma_0(\phi_{L1})]/\sigma$, which is size independent. 
The second term in Eq.~\eqref{Delta_gamma} captures the pressure-induced correction to the interfacial tension.

\begin{figure}
    \begin{minipage}{0.49\columnwidth}
        \centering
        \includegraphics[width=1.1\linewidth]{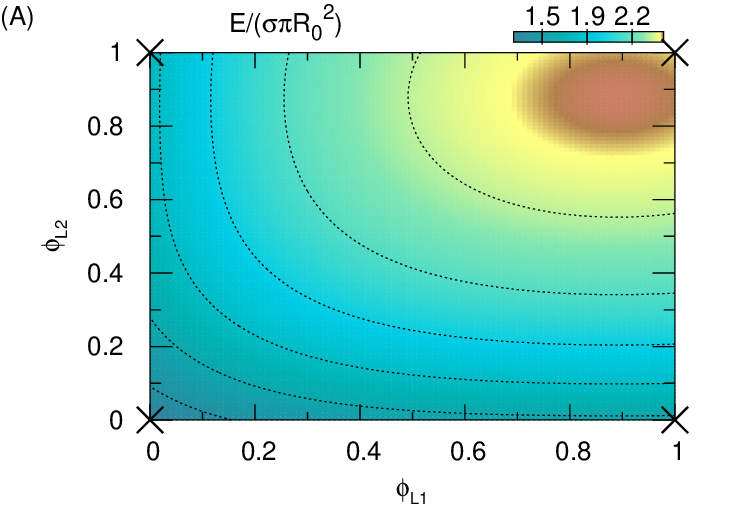}
    \end{minipage}
    \hfill
    \begin{minipage}{0.49\columnwidth}
        \centering
        \includegraphics[width=1.1\linewidth]{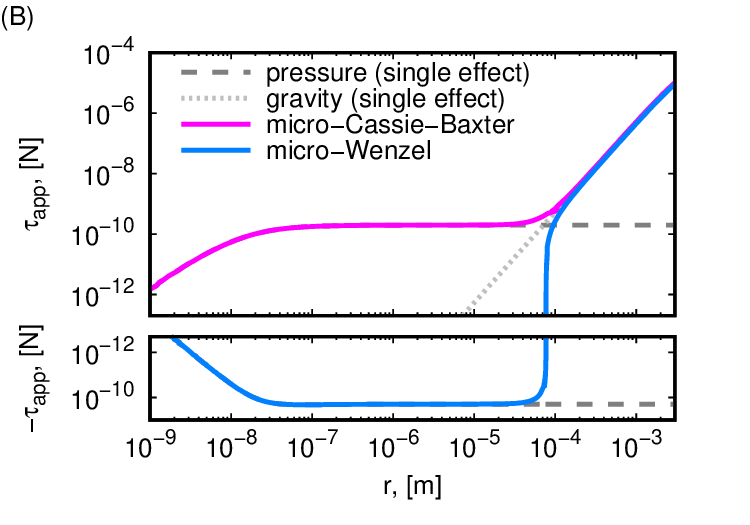}
    \end{minipage}
    \begin{minipage}{0.49\columnwidth}
        \centering
        \includegraphics[width=1.1\linewidth]{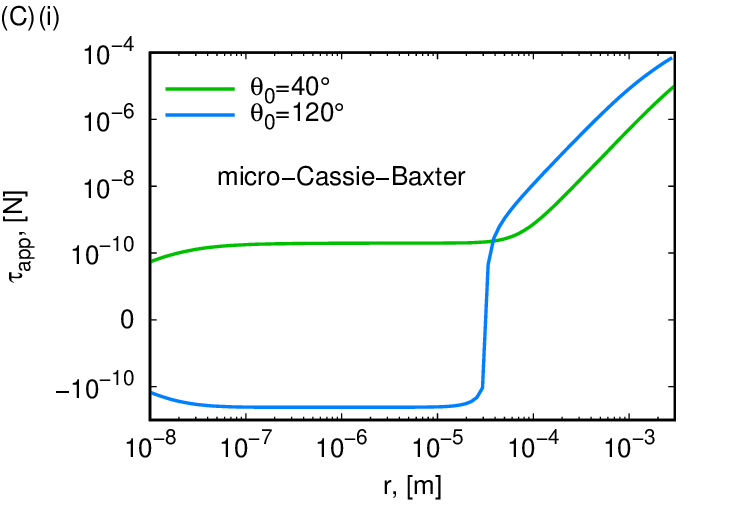}
    \end{minipage}
    \hfill
    \begin{minipage}{0.49\columnwidth}
        \centering
        \includegraphics[width=1.1\linewidth]{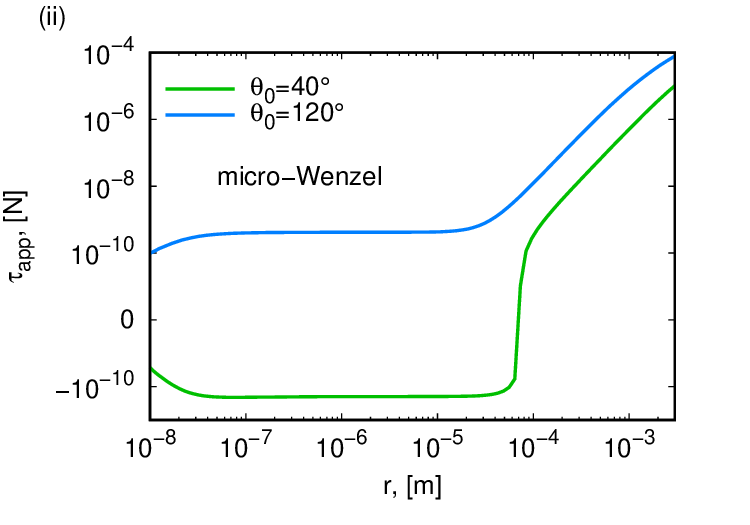}
    \end{minipage}
    \caption{(A) Energy landscape $E(\phi_{L1},\phi_{L2})$ with local energy minima marked by crosses with $S_2=0.1S_1$. (B) Line tension as a function of the base radius: single pressure effect (dashed lines), single gravity effect (dotted lines), combined effect (solid lines; micro--Cassie--Baxter state: pink, micro--Wenzel: blue). (C) Apparent line tension as a function of the base radius for different $\theta_0$ (different materials) at (i) micro--Cassie--Baxter state and (ii) micro--Wenzel state.}
    \label{Fig3}
\end{figure}

We identify the local energy minima~\cite{wangWettingContactAngleHysteresis2024,aurbachWettingPhenomenaDroplets2024} of $E(\phi_{L1},\phi_{L2})$ at $\theta$ minimizing the energy at each point, as defined by Eq.~\eqref{eq-energy}, over the domain $\Omega:=\{(\phi_{L1},\phi_{L2})\vert 0\leq\phi_{L1}\leq1,0\leq\phi_{L2}\leq1\}$ (\autoref{Fig3} (A)).
We refer to the local energy minimum at
\begin{align}
    \phi_{L2}^{CB}=1 \text{ and }\phi_{L1}^{CB}=0
\end{align}
as the micro--Cassie--Baxter state (CB), in which no liquid penetrates into the adsorption layer beneath the droplet.
There is a second local energy minimum, the micro--Wenzel state (W), located at
\begin{align}
    \phi_{L2}^W=0 \text{ and }\phi_{L1}^W=1,
\end{align}
where no gas penetrates into the adsorption layer beneath the liquid droplet.
Here, the material parameters are fixed by Young's contact angle,
\begin{align}
    \cos \theta_{0,CB/W}=\mp\frac{\ell}{\sigma}(\Delta \epsilon-\Delta \chi).
\end{align}
The remaining local energy minima at $(\phi_{L1},\phi_{L2})=(0,0)$ and $(\phi_{L1},\phi_{L2})=(1,1)$ correspond to states with $\Delta\phi=0$ and thus exhibit no pressure-induced line tension. They are therefore excluded from the following analysis. 
The coexistence of two metastable minima separated by an energy barrier directly results in contact angle hysteresis, even on ideally smooth substrates.

By applying the volume constraint
\begin{align}
    R(\theta)=\left[\frac{3V}{\pi(1-\cos\theta)^2(2+\cos \theta)}\right]^\frac{1}{3} \label{volume-constraint}
\end{align}
to Eq.~\eqref{Delta_gamma}, we obtain $\Delta \gamma$ that depends explicitly on $\theta$ and $V$. This dependence leads to a modified energy function and a modified energy-minimizing contact angle, resulting in line tension. 
Young's law is no longer applicable to these contact angles, as the additional term in Eq.~\eqref{Delta_gamma} depends on $\theta$.
Here, the pressure acts not only as a global effect but also induces a localized response within the adsorption layer.

Substituting Eq.~\eqref{Delta_gamma} and the volume constraint Eq.~\eqref{volume-constraint} into Eq.~\eqref{eq-energy}, and minimizing the energy $E(\theta)$ yields
\begin{align}
    \cos \theta_0=c&-\frac{\ell \Delta \phi}{2R(\theta)}(-c^2-2c+1)-\frac{\Delta\rho g}{6\sigma}R(\theta)^2(1-c)^2, \label{sol-theta}
\end{align}
with $c=\cos \theta$, which is solved for $\theta$ at each droplet size.
From this procedure, we obtain $\cos \theta$ as a function of $1/r$, a relationship that is frequently observed to deviate from linearity over several orders of magnitude in droplet size in experimental studies~\cite{iwamatsuGeneralizedYoungsEquation2018,heimMeasurementLineTension2013,khalkhaliWettingNanoscaleMolecular2017,scocchiWettingContactlineEffects2011,tanBodyForcesDrive2023,tatyanenkoLineTensionDualgeometry2025,checcoNonlinearDependenceContact2003}. As a result, the line tension is typically determined by the local slope, $\tau=-\sigma \frac{\mathrm{d}\cos\theta}{\mathrm{d}(1/r)}$.

\textit{Droplet size effect ---} \autoref{figures/r_tau_Zhao_mult.eps} shows the droplet size dependence of the apparent line tension at a constant Young's contact angle $\theta_0$ as a function of the base radius for both local energy minima, obtained by varying the droplet volume.
The good agreement of the theoretical curves with the experimental and simulation data points demonstrates the validity of the theory outlined above for predicting line tension.
Experimental measurements agree with the line tension associated with both the micro--Wenzel and micro--Cassie--Baxter states, despite the lower energy of the latter. 
The resulting magnitudes span from $\unit{pN}$ to $\unit{\mu N}$.
The magnitude of the line tension remains unaffected by the choice of adsorption thickness $\ell$, which is measured in a few nanometers.
The influence of different Young's contact angles on line tension is discussed below.

\autoref{Fig3} (B) shows the individual contributions to the line tension as a function of the base radius, namely the pressure effect and the gravity effect. This representation reveals the droplet-size ranges in which each contribution dominates. For small droplets ($r<\qty{3e-5}{m}$), the line tension is dominated by the pressure-induced contribution, while the gravitational effect on the line tension is negligible.
For larger droplets ($r>\qty[print-unity-mantissa = false]{1e-4}{m}$), the line tension is dominated by gravitational effects. 
An intermediate regime exhibits a continuous crossover between the two regimes.
The pressure-induced line tension remains approximately constant at $\sim \qty[print-unity-mantissa = false]{1e-10}{N}$ for $r>\qty[print-unity-mantissa = false]{1e-7}{m}$, while the gravitational line tension increases strongly with $r$.

The micro--Cassie--Baxter state and the micro--Wenzel state lead to opposite signs of $\Delta \phi$ in Eq.~\eqref{Delta_gamma}, and thus to opposite signs in the pressure-induced line tension (positive and negative, respectively). 
Because the gravitational line tension is always positive, the total line tension changes sign in the crossover regime for the micro--Wenzel state, while remaining strictly positive for the micro--Cassie--Baxter state.

\autoref{Fig3} (C) shows the line tension as a function of the base radius for different Young's contact angles $\theta_0$ at (i) the micro--Cassie--Baxter state and (ii) the micro--Wenzel state. 
For nanometric and micrometric droplets, the sign of the line tension depends on Young's contact angle. As illustrated in \autoref{Fig3} (C), the line tensions of the two states are either both positive or of opposite signs. In contrast, for millimeter-scale droplets, the line tension is dominated by gravitational effects and is invariably positive.

\begin{figure}
    \centering
    \includegraphics[width=\linewidth]{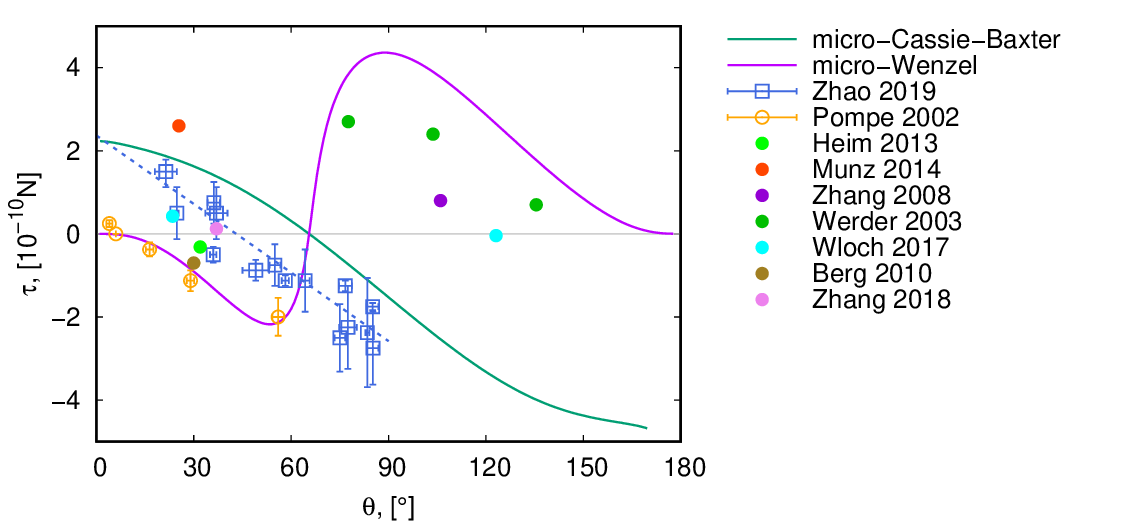}
    \caption{Apparent line tension at the micro--Cassie--Baxter and the micro--Wenzel state as a function of the apparent contact angle $\theta$ in comparison with experimental measurements~\cite{zhaoResolvingApparentLine2019,pompeLineTensionBehavior2002}.}
    \label{figures2/theta_tau_Zhao_new2.eps}
\end{figure}

\textit{Origin of the sign for microscopic droplets ---} Having established the size dependence, the subsequent analysis focuses on the role of surface wettability. \autoref{figures2/theta_tau_Zhao_new2.eps} shows the line tension as a function of the apparent contact angle~\cite{marmurLineTensionIntrinsic1997} at the micro--Cassie--Baxter and micro--Wenzel states, respectively. A linear regression of $\cos \theta$ as a function of $1/r$ was performed over the range of $\qty{16}{nm}<r<\qty{116}{nm}$, from which the line tension is obtained as $-\sigma$ times the slope of the fitted straight line. Within the range of droplet sizes considered, gravitational effects are negligible. 
The micro--Cassie--Baxter state exhibits a positive line tension at small contact angles and a negative line tension at large contact angles, with a sign change at $\theta\approx66^\circ$, whereas the micro--Wenzel state displays the opposite behavior, with the same transition angle. 
The dependence of $\Delta\gamma$ on $R(\theta)$ leads to a shift in the energy-minimizing contact angle $\theta$ due to the volume constraint. 
Neglecting bulk energy contributions, the displacement of the equilibrium angle with respect to $\theta_0$ follows from the equilibrium condition $\mathrm{d}E/\mathrm{d}\theta=0$,
\begin{align}
    \cos \theta&-\frac{\ell\Delta\phi}{R(\theta)}f(\theta)=\cos \theta_0\label{displacement}\\
    \text{with }f(\theta)&=\frac{R'(\theta)\sin\theta+2R(\theta)\cos \theta}{R'(\theta)\sin\theta+R(\theta)\cos\theta}\nonumber\\
    &=-\cos ^2\theta-2\cos \theta+1.
\end{align}
Due to a crossover of the dominant contributions from $R'(\theta)\sin \theta$ to $R\cos \theta$, $f(\theta)$ changes sign with increasing $\theta$, thereby reversing the displacement direction in Eq.~\eqref{displacement} and the sign of the line tension.
Zhao et al.~\cite{zhaoResolvingApparentLine2019} observed a sign change from positive to negative line tension at $\theta\approx43^\circ$. The experimentally determined magnitude of approximately $\qty[print-unity-mantissa = false]{1e-10}{N}$ is in good agreement with the theoretical predictions. 
By comparing both local energy minima, we find that the micro--Cassie--Baxter state is energetically favored over the micro--Wenzel state for all hydrophilic contact angles. A possible cause for the deviations from the experimental data by Zhao et al.~\cite{zhaoResolvingApparentLine2019} may be the use of ionic liquids, for which electrostatic interactions~\cite{zhangExplorationContactAngle2024} can contribute to both the contact angle and the line tension.

\textit{Macroscopic roughness effect ---} In the macroscopic Wenzel state~\cite{bormashenkoYoungBoruvkaNeumann2009,whymanRigorousDerivationYoung2008,BiWettingTransition2025,zhangExplorationContactAngle2024}, fully wetted macroscopic surface roughness rescales the solid–liquid interfacial area in Eq.~\eqref{eq-energy} by the Wenzel roughness factor $f_W$. Eq.~\eqref{displacement} then becomes
\begin{align}
\frac{\cos \theta}{f_W}-\frac{\ell \Delta\phi}{R(\theta)}f(\theta)=\cos \theta_0.
\end{align}

An analogous analysis for the macroscopic Cassie--Baxter state~\cite{bormashenkoYoungBoruvkaNeumann2009,whymanRigorousDerivationYoung2008,BiWettingTransition2025,zhangExplorationContactAngle2024}, with a solid fraction $\psi$, yields 
\begin{align}
    \frac{\cos \theta}{\psi}-\frac{\ell\Delta\phi}{R(\theta)}f(\theta)+\frac{1-\psi}{\psi}=\cos\theta_0.
\end{align}
For typical roughness factors ($1<f_W<3$~\cite{HongruMeasurement2017} and $0\ll\psi<1$), the macroscopic Wenzel and Cassie--Baxter states do not change the order of magnitude of the line tension.
Tadmor~\cite{tadmorLineEnergyRelation2004,tadmorLineEnergyLine2008} associated line tension with chemical and topological surface defects and the resulting contact angle hysteresis.
However, the classic Wenzel and Cassie--Baxter equations alone do not show size-dependent contact angles, and no defect-induced line tensions are observed, in contrast to the experimentally observed line tension at rough surfaces of order $\qty[print-unity-mantissa = false]{1e-10}{N}$~\cite{LongDistangle2025}. Including microscopic roughness resolves this discrepancy. Consequently, macroscopic roughness contributes only a subleading effect to the observed line tension.

\textit{Conclusion ---} In summary, we present a theory for the apparent line tension that combines the effect of body energy and pressure-induced contributions to the interfacial tensions within an adsorption layer.
The theory quantitatively reproduces experimental observations, including sign reversals and variations over several orders of magnitude, which cannot be captured by individual contributions alone.
We identify the adopted metastable microscopic wetting state, the surface wettability, and droplet size as the key parameters controlling the sign of the apparent line tension, with droplet size also affecting its magnitude.
Distinct physical mechanisms dominate in different droplet size regimes: pressure-induced effects at the adsorption layer for nanometric and micrometric droplets, and gravitational contributions for millimetric droplets. 
Our results indicate that the wide range of measured line tensions does not reflect inconsistencies among results but rather the diversity of the investigated systems~\cite{amirfazliStatusThreephaseLine2004}. 
This work provides a physical interpretation of the sign and magnitude of line tension across droplet sizes and experimental systems.

\begin{acknowledgments}
F.A. thanks the Gottfried-Wilhelm Leibniz prize NE 822/31-1 of the German Research Foundation (DFG) for funding this research. 
F.W. is grateful to the VirtMat project P09 
of the Helmholtz Association (MSE-programme No. 43.31.01).
\end{acknowledgments}
\appendix
\section*{END MATTER}
\textit{Interfacial tensions ---}
The liquid-solid and solid-gas interfacial tensions consist of three terms:
\begin{align*}
    \gamma(\phi_L,\phi_G)=\ell [\epsilon(\phi_L,\phi_G)+p(\phi_L,\phi_G)+\chi(\phi_L,\phi_G)].
\end{align*}
The volume fraction of the liquid and gas in the adsorption layer is denoted by $\phi_L$ and $\phi_G$, respectively. 
The internal energy is represented by $\epsilon=\epsilon_L\phi_L+\epsilon_G\phi_G$, where $\epsilon_L$ and $\epsilon_G$ are the internal energies of the liquid and gas phases, respectively. Furthermore, the van der Waals energy is denoted by $\chi=\chi_{LG}\phi_L\phi_G+\chi_{SG}\phi_G+\chi_{SL}\phi_L$, with the van der Waals interactions between the liquid-gas, solid-gas, and solid-liquid phases $\chi_{LG}$, $\chi_{SG}$, and $\chi_{SL}$, respectively. The pressure term is expressed as $p=p_L\phi_L+p_G\phi_G$.
Inserting the constraint $\phi_L+\phi_G=1$ yields a quadratic polynomial in $\phi_L$
\begin{align*}
    \gamma(\phi_L)&=\gamma(\phi_L,1-\phi_L)\\&=\ell \chi_{LG} \left[-\phi_L^2+\left(1-\frac{\Delta \epsilon+\Delta p-\Delta \chi}{\chi_{LG}}\right)\phi_L+\xi \right],
\end{align*}
with $\Delta \epsilon=\epsilon_G-\epsilon_L$, $\Delta p=p_G-p_L$, $\Delta \chi=\chi_{SL}-\chi_{SG}$, and $\xi=\frac{\epsilon_G+p_G+\chi_{SG}}{\chi_{LG}}$. 

\textit{Local droplet penetration area ---}
The local droplet penetration area $S_2$ describes the area around the droplet at the solid-liquid interface, where the liquid penetrates into the solid at the adsorption layer and modifies the interfacial tension by the liquid volume fraction. It is considered to be a fraction of the wetted area $S_1$,
\begin{align*}
    S_2=\alpha S_1,
\end{align*}
with $0<\alpha\ll1$. This factor influences the number and location of the local energy minima of $E(\phi_{L1},\phi_{L2})$ (see \autoref{Fig3} (A)) \cite{aurbachWettingPhenomenaDroplets2024}.

\sloppy
\bibliography{bibliography}

\end{document}